\documentclass{sig-alternate}
\usepackage[T1]{fontenc}
\usepackage{cite}
\usepackage{xspace}
\usepackage{url}
\usepackage{epsfig}
\usepackage{color}
\usepackage{times}
\usepackage{pgf}
\usepackage{tikz}
\usepackage{alltt}
\usepackage{booktabs}
\usepackage{tabularx}
\usepackage{enumitem}
\usepackage{comment}
\usepackage{textgreek}
\usepackage{paralist}
\usepackage[textfont=bf,labelfont=bf]{caption}
\usepackage{graphicx}
\usepackage{subfigure}
\usepackage{etoolbox}

\makeatletter
\def\@copyrightspace{\relax}
\makeatother

\newcommand{\figdir}[1]{#1}

\renewenvironment{itemize}{
   \begin{list}{\labelitemi}{
     \setlength{\topsep}{0.5ex}
     \setlength{\itemsep}{-0pt}
     \setlength{\itemindent}{0pt}
     \setlength{\leftmargin}{\labelwidth}
     \addtolength{\leftmargin}{-8pt}}
}{\end{list}}

\newcommand{\locking}{\textit{Deadlock free locking}}
\newcommand{\splitSystemName}{\textsc{Split ORTHRUS}}
\newcommand{\splitLocking}{\textit{Split Deadlock free}}
\newcommand{\systemName}{\textsc{Orthrus}}
\newcommand{\partitionedStore}{\textit{Partitioned-store}}

\newcommand{\noeditingmarks}{}

\newcommand{\textred}[1]{\textcolor{red}{#1}}

\ifx\noeditingmarks\undefined
   \newcommand{\pgwrapper}[2]{\textbf{#1: }\textred{\textit{#2}}}
   
\else
   \newcommand{\pgwrapper}[2]{}
   
\fi

\newcommand{\jose}[1]{\pgwrapper{Jose}{#1}}
\newcommand{\dna}[1]{\pgwrapper{DNA}{#1}}

\newcommand{\secref}[1]{Section~\ref{sec:#1}}
\newcommand{\figref}[1]{Figure~\ref{fig:#1}}

\newcommand{\etal}{et al.\@}

\usepackage{soul}

\begin{document}

\twocolumn
\newpage

\setlength{\belowcaptionskip}{-14pt}
\title{
  Design Principles for Scaling Multi-core OLTP Under \\High Contention{\titlenote{This is a draft of 
work that is accepted to appear at SIGMOD 2016.}}
}

 \numberofauthors{3}
 \author{
 \alignauthor
 Kun Ren\\
 \affaddr{Yale University}\\
 \email{kun.ren@yale.edu}\\
 \alignauthor
 Jose M. Faleiro\\
  \affaddr{Yale University}\\
  \email{jose.faleiro@yale.edu}
  \alignauthor
  Daniel J. Abadi\\
  \affaddr{Yale University}\\
  \email{dna@cs.yale.edu}
  }

\maketitle

\begin{abstract}
%

Although significant recent progress has been made in improving the
multi-core scalability of high throughput transactional database
systems, modern systems still fail to achieve scalable throughput for
workloads involving frequent access to highly contended data. Most of
this inability to achieve high throughput is explained by the
fundamental constraints involved in guaranteeing ACID --- the addition
of cores results in more concurrent transactions accessing the same
contended data for which access must be serialized in order to
guarantee isolation. Thus, linear scalability for contended workloads
is impossible. However, there exist flaws in many modern architectures
that exacerbate their poor scalability, and result in throughput that
is much worse than fundamentally required by the workload.

In this paper we identify two prevalent design principles that limit
the multi-core scalability of many (but not all) transactional
database systems on contended workloads: the multi-purpose nature of
execution threads in these systems, and the lack of advanced planning
of data access. We demonstrate the deleterious results of these design
principles by implementing a prototype system, \systemName{}, that is
motivated by the principles of separation of database component
functionality and advanced planning of transactions. We find that
these two principles alone result in significantly improved
scalability on high-contention workloads, and an order of magnitude
increase in throughput for a non-trivial subset of these contended
workloads.

%

\end{abstract}
\section{Introduction}
\label{sec:intro}
The maximum throughput that a database system is able to achieve is
dependent on many factors, from the hardware on which it runs to the
particular implementation details of the database software. While most of
these factors can be overcome by spending more money, on better
hardware or better software developers, throughput will always be
fundamentally limited by the presence of {\it contended operations} in a
workload. 

The definition of a ``contended operation'' may vary depending on the
user's requested isolation level of transactions, the ability of the
database to prevent reads from conflicting with writes via
multi-versioning, and the semantic commutativity of
operations. Nonetheless, unless no isolation whatsoever is required,
there will always be certain operations that cannot be executed
concurrently, and the presence of many of these contended operations
in a workload will necessarily limit throughput. Thus, adding more
processors to a system, which enables more transactions to be
processed in parallel, only increases throughput if the additional
transactions that can be run do not conflict with the existing
transactions that are currently running.


For decades, database systems had been designed for single processor
machines.  These conventional database architectures were ill suited
for the abundant parallelism in multi-core hardware. As a consequence,
they could not achieve scalable throughput across the entire spectrum
of transactional workloads. Particularly problematic was the fact that
conventional database architectures were not able to scale throughput
on {\it low contention} workloads, despite the fact that low
contention workloads have no fundamental limit to scalability.  To
address this gap between hardware and database software, most work on
multi-core database systems has focused on eliminating fundamental
scalability bottlenecks in these systems' design
\cite{johnson2009improving,jung2013scalable,tu2013speedy}.  As a
result of this research, today's state-of-the-art systems can achieve
close to linear scalability in transactional throughput on low
contention workloads.

Unfortunately, as recently demonstrated by Yu et al., multi-core
database systems continue to be plagued by performance problems on
{\it high contention} workloads \cite{yu2014staring}.  For the
fundamental reason described above, it is impossible to achieve linear
scalability in high contention workloads; the throughput of a database
system should taper as cores are added under high contention. 
However, the \textit{actual} shortfall relative to linear
scalability on high contention workloads is much worse than what is
theoretically required by the nature of the contention in the
workload. In some cases, in fact, throughput actually {\it decreases}
as more cores are added. The problem is that the overhead of managing
transactions, particularly that of concurrency control, is
proportional to the amount of workload contention. At high levels of
contention, concurrency control overhead takes up a non-trivial
fraction of each transaction's total execution time.  As a
consequence, modern multi-core database systems achieve nowhere near
the theoretical performance determined by the achievable concurrency
in high contention workloads.

%

We attribute this poor performance under high contention to two design
decisions that, to the best of our knowledge, are present in every
widely-used transactional database system available today. First,
despite compelling proposals to the contrary
\cite{stageddb,pandis2010data,pandis2011plp}, database systems tend to
assign responsibility for a particular transaction to a single thread
\cite{hellerstein2007architecture}.  Assigning a transaction to a
single thread \textit{conflates database functionality}, which leads
to poor instruction and data cache locality \cite{stageddb}. Worse,
this conflated functionality causes workload contention to directly
impact physical contention in the database system (\secref{conflated}).

Second, database systems allow dynamic access of data without advanced
planning of transactions' data access patterns. The negative
side-effects of this design decision is most clearly present in
database systems that use pessimistic concurrency control protocols
based on logical locking.  Such systems dynamically acquire each
transaction's logical locks in an arbitrary order, which makes
transactions susceptible to deadlocks. Any system which employs
dynamic lock acquisition must therefore include a mechanism for
handling deadlocks. Under high contention workloads, deadlock handling
mechanisms are a significant source of overhead, and can lead to
wasted work due to transaction aborts (\secref{interactive}).


This paper proposes two design principles to address these overheads.
First, we propose that database systems \textit{partition
functionality} across the cores of a single machine. Instead of using
a single thread to perform both a transaction's logic \textit{and}
concurrency control on behalf of the transaction, we dedicate a set of
threads to perform \textit{only} concurrency control, and another set
of threads to execute transaction logic. Concurrency control and
execution threads cooperatively process transactions using
explicit message-passing. 

Second, we propose that database systems analyze transactions prior to
their execution in order to predict their access patterns. These
access patterns are used to coordinate access to data. Since
almost all widely-used database systems use pessimistic locking to
protect at least some types of data access, our focus in this paper is
particularly within the context of pessimistic locking protocols. In
this context, planning data access enables the implementation of a
deadlock avoidance protocol, which circumvents the overhead of
deadlock detection and resolution.

We built a prototype database system, \systemName{}, based on the
design principles of partitioned functionality and deadlock freedom. A
notable aspect of \systemName{}'s design is the use of message passing
between cores devoted to different functionalities.  Although the use
of explicit message-passing among a system's cores has been used in
the context of multi-core operating systems
\cite{baumann2009multikernel,wentzlaff2009factored} and programming
models \cite{calciu2012delegation,lozi2012remote}, to the best of our
knowledge, \systemName{} is the first multi-core database system to
successfully use explicit message-passing to avoid physical contention
on shared concurrency control data structures in the database system.

To summarize, the contributions of this paper are as follows:
\begin{itemize}

\item We identify two sources of overhead in state-of-the-art
  multi-core database systems that severely limit throughput under
  high contention workloads: conflated functionality
  (\secref{conflated}) and dynamic data access (\secref{interactive}).

\item We propose two design principles to address this overhead;
partitioning database functionality (\secref{cc}) and planned data
access (\secref{avoidance}).

\item Based on these design principles, we implement a prototype
database system, \systemName{}. We discuss techniques to make our
design principles practical to implement (\secref{optimizations} and
\secref{alternative}).

\item We perform an extensive set of experiments in order to evaluate
the multi-core scalability of \systemName{} relative to an archetypal
modern multi-core database system (\secref{eval}).

\end{itemize}

\section{Problems with Existing Designs}
\label{sec:bg}

\subsection{Conflated functionality}
\label{sec:conflated}
As mentioned above, in most database systems, a single thread
processes an individual transaction
\cite{hellerstein2007architecture}.  This single thread manages both
the execution of the transaction's logic and the necessary
interactions with the concurrency control module of the database
system (e.g. a lock manager or the shared data structures needed for
OCC validation).  Several such threads concurrently execute on a
single multi-core system.  These concurrently executing threads make
requests of the same concurrency control module. This design pattern
of multiple threads globally sharing a single concurrency control
module can lead to severe scalability bottlenecks. We discuss these
bottlenecks in this section. 


\textbf{Synchronization overhead.}  
We first discuss the overhead associated with synchronization on
concurrency control meta-data. 
In order to control the
interleaving of concurrent transactions, any concurrency control
protocol must associate some meta-data with the database's logical
entities.  The meta-data used is protocol dependent. For example,
pessimistic locking protocols use a hash-table of lock requests on
records \cite{gray1993transaction}, while optimistic and multi-version
protocols associate timestamps with records
\cite{larson2011concurrency,tu2013speedy}. As part of the concurrency
control protocol, several concurrent threads may need to read or write
the meta-data associated with a particular database object. 
Concurrent threads must therefore synchronize their access to concurrency
control meta-data.  Note that synchronization is \textit{not}
implementation dependent; instead, it is intrinsic to any concurrency
control protocol whose meta-data can be read or written by any
database thread.  Since meta-data is associated with database objects
(such as records), contention for concurrency control
meta-data is directly affected by workload contention; if a particular
database record is popular, then threads will need to frequently
synchronize on that record's meta-data. Unfortunately, on modern
multi-core hardware, contention
significantly degrades the performance of \textit{atomic instructions}
\cite{boyd-wickizer2012non-scalable,faleiro2015multiversion,tu2013speedy}.
These atomic instructions -- such as \textit{fetch-and-increment} and
\textit{compare-and-swap} -- are the basic building blocks of both
latch based and latch-free algorithms.  Thus, under contention, both
classes of algorithms are susceptible to severe performance
degradation.

\textbf{Data movement overhead.}
In addition to synchronization on concurrency control meta-data,
another source of overhead in conventional database architectures is the
\textit{movement} of this meta-data across multiple cores.
In order to access an object's meta-data, a thread must move
the memory words corresponding to the meta-data into its CPU core's
local cache.  As multiple threads request access to a particular
object's meta-data, the memory words corresponding to the meta-data
are moved between cores.  The movement of data between a machine's
cores occurs because multiple threads are allowed to read or write
the data. 
If a thread requires access to a
latch-protected data-structure, the thread must first acquire the
latch, and then move the data-structure to its core. 
Only when these two steps complete can the thread actually access the
data-structure. Since the latch is acquired first, it is held for the
time it takes to move the data-structure. As a consequence,
data-movement extends the duration for which latches are held. 
In the presence of contention, the increase in latch hold times
can contribute to a decrease in concurrency.

  \begin{figure}
  \centering
  \includegraphics[width=\linewidth]{\figdir{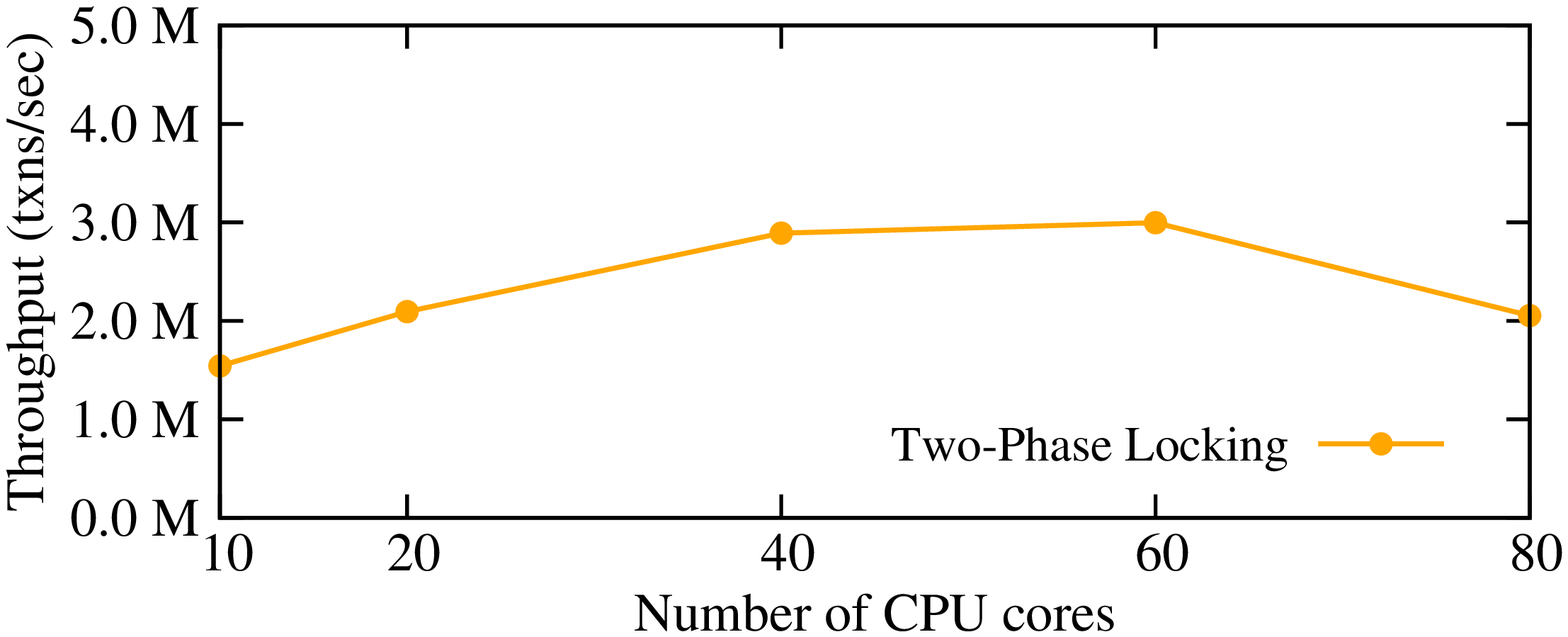}}
  \caption{Scalability of read-only transactions under two-phase locking
    on a high contention workload.}
  \label{fig:scalability_bg}
  \end{figure}

In order to validate the deleterious effects of synchronization and
data movement overhead, we ran a simple experiment to measure the
scalability of short read-only transactions under two-phase locking on
a high contention workload (see \secref{ycsb_eval} for a detailed
description of the experimental setup).  Since transactions are
read-only, the workload is conflict free (despite the presence of
contention). \figref{scalability_bg} shows the results of the
experiment. \figref{scalability_bg} shows that two-phase locking is
unable to scale beyond 40 cores despite the absence of conflicts, and
surprisingly \textit{decreases} in performance. The inability to scale
is due to synchronization and data movement overhead. The source of
synchronization overhead is two-phase locking's use of atomic
instructions on contended memory words to manipulate the table of lock
requests.  Data movement overhead is a consequence of multiple cores
manipulating the same list of lock requests (due to multiple cores
requesting read locks on the same records).

%

\textbf{Instruction and data cache pollution.}
If a single thread executes both concurrency control and transaction
logic, then the thread's CPU core must cache data and instructions
corresponding to each of these two functions. 
The data and instructions corresponding to concurrency control and
transaction logic thus compete for a single core's cache. Concurrency
control cache-lines therefore evicts transaction logic cache-lines,
and vice-versa. This has the overall effect of increasing the duration
of each transaction, which in turn decreases overall throughput. 




All three reasons for performance degradation --- synchronization
overhead, data movement overhead, and cache pollution --- have the
effect of increasing a transaction's execution time. Not only does
this increase in execution time per transaction necessarily reduce
throughput by occupying system resources for longer periods of time,
but the throughput reduction is compounded by the fact that increasing
transaction time is particularly harmful in high contention
settings. This is because conflicting transactions either have to
block (in pessimistic schemes) or abort (in optimistic schemes). The
longer it takes to execute a transaction, the higher the probability
that a later conflicting transaction will abort or block
behind the original transaction.

Note that database systems such as H-Store \cite{hstore} and Hyper
\cite{kemper11hyper} do not suffer from the overheads associated with
conflated functionality. These systems employ coarse-grained
partition-level concurrency control, eliminating synchronization and
data-movement overhead for single-partition transactions. 
Furthermore, coarse-grained
partition-level concurrency control algorithms and meta-data are
simpler and significantly smaller, respectively,  
than fine-grained concurrency control. As a
consequence, these systems do not suffer from significant
cache pollution.  
However, these systems's performance degrades in the
presence of multi-partition transactions; transactions which need to
access data stored in multiple partitions.




\subsection{Dynamic concurrency control}
\label{sec:interactive}

Most database systems allow transactions to dynamically request
records to read and write as they execute. This lack of advanced
planning precludes opportunities to coordinate access to contended
items in order to maximize concurrency. The clearest example of this
are in systems that use two-phase locking (2PL) for concurrency
control, where locks are acquired for a transaction as
each data access request for a transaction is processed. 
Allowing transactions to dynamically request access to records
necessitates dynamic lock acquisition. arbitrary process of acquiring
multiple locks in an arbitrary order can lead to deadlock. These
systems must therefore implement mechanisms to deal with deadlocks. 

%

Under high contention workloads, deadlock handling logic is a significant source
of overhead.  There are two reasons for this overhead; first, deadlock
handling logic extends the duration for which locks are held, second,
deadlocks waste useful work performed by transactions that must be
aborted. 

\begin{itemize}

\item
Since a transaction cannot deadlock unless it has already acquired
locks, any deadlock handling logic must be executed while locks are
held. Therefore, deadlock handling logic extends the duration for
which locks are held. The increased lock hold time means that
conflicting transactions must wait longer to execute.
Therefore, deadlock handling logic imposes a
performance penalty regardless of whether a deadlock has actually
occurred.

\item
If deadlock handling logic aborts a transaction, then any work
performed by the transaction is wasted. Furthermore, if a transaction
is allowed to directly write records (i.e., if transactions do not
buffer their writes), then the database must also undo the aborted
transaction's writes. In addition to wasted work, aborted transactions
induce unnecessary waiting on conflicting transactions; if a
conflicting transaction is made to wait on a transaction that is
eventually aborted, then in retrospect the conflicting transaction
could have been allowed to make progress without delay. Finally, under
high contention workloads, deadlocks are simply more
prevalent. Therefore, the wasted work and unnecessary waiting due to
transaction aborts occur more often under high contention workloads.
\end{itemize}

\section{Architecture}
\label{sec:arch}
In this section, we describe the architecture of a proof-of-concept
system that we built, \systemName{}. \systemName{} is designed to
ameliorate the scalability bottlenecks described in \secref{bg}.
\systemName{} is not a complete database system --- we did not build a
relational query processor, a client communications manager, or many
of the shared utilities that are present in most database
systems. Instead, we just built the transaction management component
of the system, with a particular focus on concurrency control, since
the main impediment to scalability under high contention is
concurrency control.

\systemName{} implements locking-based pessimistic concurrency
control. \systemName{} is targeted at workloads with high contention,
and optimistic schemes are well-known to perform poorly under high
contention due to excessive aborts --- even recent proposals for
multi-core optimized optimistic schemes (such as Silo
\cite{tu2013speedy}and Hekaton \cite{larson2011concurrency}) perform
poorly under high contention \cite{faleiro2015multiversion}.

Like most recently proposed architectures for transactional database
systems, \systemName{} assumes that the working set of data accessed
by transactions can be held in main memory, since memory sizes are
growing faster than transactional working sets \cite{hstore}. As a
consequence, \systemName{} does not incur stalls to access data from
disk.  \systemName{} therefore creates exactly the same number of
threads as physical CPU cores, and pins each thread to a single core, as is
done in several other main-memory database systems
\cite{faleiro14lazy,faleiro2015multiversion, hstore,pandis2010data,pandis2011plp,ren13lightweight,tu2013speedy,kemper11hyper,yu2014staring,narula2014phase}.



\systemName{}'s first key design feature is that it \textit{partitions
functionality} across the cores of a single machine. The thread pinned
to each core on the server is devoted to a single narrow component of
transaction processing. Since our focus in this paper is on
concurrency control, \systemName{} assigns cores one of two possible
roles; concurrency control or transaction execution. Note, however,
that this philosophy can extend to other roles within a database system. For
instance, StagedDB successfully separated functionality across cores in
the query processor component of the database system \cite{stageddb}. 
In \systemName{}, concurrency control and transaction execution cores do not
share any data-structures; concurrency control cores cannot read or
write any data-structures on execution cores, and vice-versa. 
Instead of implicitly communicating through shared-memory,
\systemName{} uses explicit
message-passing between concurrency control and execution threads.

\jose{The subsections that follow deal with the benefits of
  partitioned functionality. We could get rid of this paragraph.}
\dna{I deleted it but want to think harder about it later}

The second key design feature of \systemName{} is data access planning
for the purpose of deadlock avoidance. In the rest of this section, we
discuss these two key features of \systemName{}'s architecture in
more detail.

\subsection{Partitioned functionality}
\label{sec:cc}

Logically, concurrency control threads perform the same function
as that of a centralized lock manager. \systemName{} partitions
responsibility for database objects across concurrency
control threads such that each database object is controlled by single
thread.
Thus, each concurrency control thread
maintains meta-data for a disjoint subset of the database's
objects. The meta-data on each concurrency control thread is exactly the
same as in a centralized lock manager; each thread maintains a
hash-table which maps keys to a list of transaction lock requests.  

Execution threads do not contain instructions nor data pertaining
to concurrency control; they are only responsible for performing each
transaction's logic. Each transaction is assigned to a single
execution thread, which is responsible for processing the
transaction's logic in its entirety.  
Execution threads request locks by sending messages to the
concurrency control threads responsible for those locks\footnote{
Since responsibility for objects is disjointedly
  partitioned across concurrency control threads, the set of locks
  required by a single transaction may reside on multiple concurrency
  control threads.}. 
Message passing between execution threads and concurrency control
threads is mediated 
via queues. Space is allocated in shared memory for an input queue to
each concurrency control thread and an output queue from each
concurrency control thread. Messages are passed via reads and writes to these
queues. 

Note that a na\"{i}ve implementation of these queues leads to the same
type of synchronization bottlenecks observed in traditional
systems. If every transaction execution thread is allowed to write to
a single input queue (for a particular concurrency control thread), then
synchronization overhead will prevent a scalable implementation of
message passing. To overcome this pitfall, our implementation assigns
each concurrency control thread a \textit{separate} queue for each
execution thread. Thus, while we mention a single \textit{logical}
input queue to each concurrency control thread, its implementation
consists of $N$ physical queues, where $N$ is the number of execution
threads.

With the above implementation, each queue has only one writer (the
associated execution thread) and one reader (the associated
concurrency control thread). The queue can be therefore implemented
using a standard latch-free circular buffer \cite{lozi2012remote} to
avoid synchronization between the reader and writer except in the rare
case where the queue fills up. Consequently, our message passing
implementation does not suffer from the synchronization costs that
contended writes to shared-memory usually encounter.

Concurrency control threads serially process requests from their
logical input queues.
On dequeueing a request for locks from its input queue, a concurrency
control thread checks the requested set of locks, and determines which
of these locks are requested on objects in its local partition. 
For each lock on an object in its logical partition, the concurrency
control inserts a lock request into its local hash-table. The
concurrency control thread responds to an acquisition request only
after it has granted all locks from the request of an individual
transaction. Note that the response may take a while; the lock
acquisition request may have to wait for prior conflicting requests to
release locks (just as in conventional locking protocols).  Instead of
waiting for responses from concurrency control threads, execution
threads begin working on other transactions. 
The interaction between
execution and concurrency control threads during lock release is similar; the
only difference is that concurrency control threads respond
immediately because lock release requests are satisfied immediately.


\begin{figure}
\centering
\includegraphics[clip,width=0.8\linewidth]{\figdir{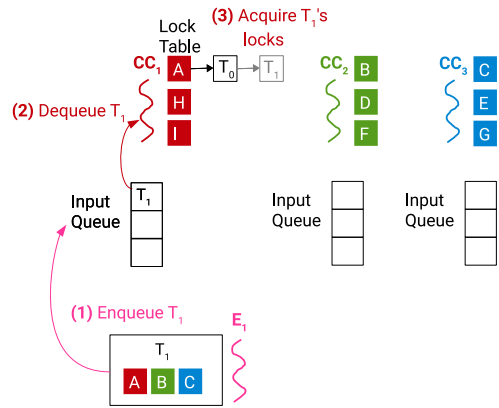}}
\caption{Concurrency control thread acquiring locks on behalf of
an  execution thread. }
\label{fig:single_lock}
\end{figure}

\figref{single_lock} shows an example of the interaction between a
concurrency control thread and an execution thread.  Transaction
$T_1$'s execution thread, $E_1$, requests locks on records $A$, $B$,
and $C$ (for simplicity, we omit the details of the \textit{lock mode}
required on each of these records).  $E_1$ writes these lock requests
in a message that we label with name of the transaction ($T_1$). In
order to acquire a lock on $A$, $E_1$ enqueues message $T_1$ in
$CC_1$'s input queue (\textbf{Step 1}).  $CC_1$ dequeues $T_1$ from its input
queue (\textbf{Step 2}), and checks the locks requested by $T_1$.  In this
particular example, of the locks requested by $T_1$ ($A$, $B$, and
$C$), only $A$ resides on $CC_1$'s partition. $CC_1$ thus inserts a
lock request on record $A$ into its local lock table (\textbf{Step 3}). The
details of how locks $B$ and $C$ are acquired, and how these requests
end up in the input queues of the other concurrency control threads
will be explained in \secref{optimizations}.

%

The communication pattern between execution and concurrency control threads
is much like client-server communication in distributed systems;
execution threads behave as clients and concurrency control threads behave as
servers. An execution thread's ``request'' is an explicit message
asking the concurrency control thread to acquire or release a lock on behalf
of a particular transaction.  While centralized lock managers and
\systemName{}'s concurrency control threads perform the same logical
function (acquire and release logical locks), they each use two
fundamentally different communication mechanisms. In conventional
database systems, threads use the abstraction of a shared-address
space to implicitly share data. Intuitively, data is shared \textit{by
default}. Since data is globally shared, each database thread can
directly manipulate the data. In contrast, in \systemName{},
the default mode of operation is to \textit{eliminate} shared data
among threads of different types; execution and concurrency control
threads do not share any data. In order to acquire and release locks
on database objects, an execution thread must request concurrency
control threads to acquire locks on its behalf.

As a result of its physical separation of concurrency control and
transaction execution concerns across threads, \systemName{}
guarantees that data is \textit{never} implicitly shared across
concurrency control and execution threads; the only way to share data
is via message-passing. The consequence of this design
principle is that \systemName{} completely eliminates all overheads
associated with conflated functionality (\secref{conflated}):

\begin{itemize}
\item
\noindent \textbf{Synchronization overhead.} \systemName{} partitions database
objects across concurrency control threads. As a consequence, every lock
acquisition and release request for a particular object is serviced by
a single concurrency control thread; reads and writes of an object's
meta-data are restricted to one thread. Therefore, this thread does not need to
synchronize its access to any of the objects's meta-data it is
responsible for. If multiple execution threads concurrently request
locks on the same record, then the requests are handled by the same
concurrency control thread. 

\item
\noindent \textbf{Data movement overhead.} Since the locking operations on a
particular database object are performed by a single thread,
concurrency control meta-data (linked-lists of lock requests in the
lock table) never need to move between threads. As a result,
\systemName{} does not suffer from any data movement overhead (for
concurrency control).

\item
\noindent \textbf{Instruction and data cache pollution.} 
In \systemName{}, execution and concurrency control threads perform two
different functions. Execution threads are completely isolated from
concurrency control threads; these two types of threads do not share any
instructions nor data. As a consequence, \systemName{} avoids the cache
pollution that is inherent in conventional database implementations.

\end{itemize}

%
%
%
%

\subsection{Deadlock avoidance}
\label{sec:avoidance}
\secref{interactive} discussed the various overheads associated with
handling deadlocks in pessimistic locking-based concurrency control
protocols. The fundamental problem behind the cost of
deadlocks is the dynamic nature of data access which stems from a lack of
advanced planning. Motivated by this observation, \systemName{} plans
data access prior to transaction execution, and leverages this
advanced planning to perform a deadlock avoidance protocol instead of
deadlock detection and resolution. 


In more detail, \systemName{} guarantees that deadlocks never occur by
enforcing a locking discipline on execution threads. Execution threads
must acquire locks on behalf of transactions in some well-defined
order.  To enforce this locking discipline, an execution thread cannot
start to make lock requests (by sending messages to concurrency
control threads) until it knows the complete set of locks requests
that it will make for a particular transaction. Once an execution
thread knows the complete set of locks requests, it can request locks
from the appropriate concurrency control threads in order of the
threads's unique identifiers. In order to avoid deadlock, these
requests to concurrency control threads cannot occur concurrently ---
locks are requested from the next concurrency control thread after the
locks from the previous thread have been granted.

In some cases, the set of locks that a transaction will request cannot
be determined via a simple inspection of the transaction
logic. Rather, there is a data-dependent access such that a part of
the transaction needs to be executed before it can be determined what
locks will be requested. In such a situation \systemName{} uses the
OLLP technique proposed by Thomson \etal{} in the context of Calvin
\cite{thomson2012calvin}. In OLLP, a transaction is partially executed
in ``reconnaissance'' mode in order to generate an estimate of the
access footprint of the transaction. No locks are acquired during this
reconnaissance, so no writes to the database state occur, and all
reads are not assumed to be consistent (this is why the resulting
access footprint is just an estimate and not a guarantee). This access
estimate is then annotated as part of the transaction, which is
submitted to the system to be actually run. Before starting to process
a transaction, \systemName{} acquires the locks that are indicated by
the access estimate annotation. If, over the course of processing the
transaction, an execution thread notices that it needs to access a
record that it did not acquire a lock for at the beginning of the
transaction (that is, the access estimate it received was incorrect),
\systemName{} updates the annotation, and then aborts and restarts the
transaction with a new estimate. Ren \etal{} have shown that the extra
overhead of OLLP (the reconnaissance phase) is generally a small
percentage of actual transaction processing, and that aborts due to
incorrect access estimates are rare in practice
\cite{calvin-experiments}. 



A disadvantage of \systemName{}'s approach relative to dynamic lock
acquisition is that there is a risk of locks being held for a longer
period of time. This is because \systemName{} immediately acquires all
locks at the beginning of a transaction, while dynamic locking can
acquire one lock at a time, interleaving lock acquisition with
transaction execution. For example, if a lock on a highly contended
record is only needed for the last operation of a transaction, 
dynamic lock acquisition only needs to hold the lock for a
very short period of time (just the end of the transaction). Meanwhile
\systemName{} must hold the lock for the entire transaction.

Note that increased lock hold time is not a disadvantage under low
contention. Increased lock hold time only hurts performance under high
contention workloads. However, it is precisely under high contention
that deadlocks are more frequent, and the cost of deadlock detection
increases (\secref{interactive}). We hypothesize that \systemName{}'s
increased lock hold time is more than offset by avoiding the decrease
in concurrency and wasted work due to deadlock detection and
aborts, respectively, in dynamic locking. We experiment with this
hypothesis in \secref{eval}.

\subsection{Optimizations}
\label{sec:optimizations}
\systemName{}'s design philosophy of partitioning concurrency control
and execution functionality across the threads of a database server
addresses several sources of overhead in conventional database
systems. Concurrency control and execution threads communicate with
each other via explicit message-passing, which is its own sources of
overhead.
The single biggest source of overhead we had to overcome was that of
\textit{asynchrony} in the interactions between execution and concurrency
control threads.

In \systemName{}, concurrency control threads run a tight loop which
sequentially processes requests for lock acquisition or release. At
any given point in time, a concurrency control thread may have
multiple outstanding lock acquisition or release requests. 
Thus, execution threads's requests may experience queueing delay
before being processed. To prevent these queueing delays from wasting
CPU cycles, execution threads do not synchronously wait on responses
from concurrency control threads. After sending a request to a
concurrency control thread, an execution thread checks whether any
older requests have received responses. If yes, the execution thread
resumes the execution of the corresponding transaction. If not, the
execution thread begins executing a new transaction.
Consequently, when a concurrency control thread does eventually
respond to a lock request, the execution thread will likely be in the
middle of working on a different transaction; responses also
experience queueing delay on execution threads.  Execution threads's
lock acquisition and release requests are therefore \textit{asynchronous}.


On its own, asynchrony is not a source of overhead. However, the
interaction of asynchrony with the higher level concept of logical
locking can hamper concurrency.  The queueing delays experienced by
requests and responses extend the duration for which logical locks are
held. For instance, if an execution thread requests a lock on record
$A$ for transaction $T$, the time between when the lock is acquired by
the appropriate concurrency control thread, and the time the execution
thread resumes the execution of $T$ represents time for which the lock
on $A$ is needlessly held. Furthermore, the overhead due to queueing
delay is compounded when an execution thread requests locks from
multiple concurrency control threads. This is because our deadlock
avoidance mechanism requires that lock requests to these concurrency
control threads occur sequentially (as explained in
\secref{avoidance}).  For instance, if the execution thread from the
example above subsequently requests locks on $B$ and $C$, the lock on
$A$ is held while requests and responses on $B$ and $C$ experience
queueing delays.

\begin{figure}
\centering
\includegraphics[clip,width=0.6\linewidth]{\figdir{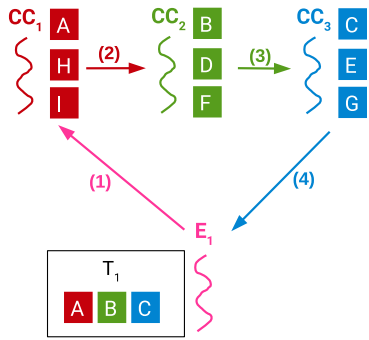}}
\caption{Concurrency control thread acquiring locks on behalf of
an  execution thread. }
\label{fig:multiple_locks}
\end{figure}

In order to reduce the impact of queueing delay, \systemName{}
minimizes the number of messages sent between threads.  The basic idea
is to make concurrency control threads to forward lock request
messages to other concurrency control threads on behalf of an
execution thread. In other words, instead of paying two message delays
for all concurrency control threads involved in a transaction, each
concurrency control thread forwards the request directly to the next
thread involved in that transaction (which reduces the number of
message delays per concurrency control thread to one).

\figref{multiple_locks} shows an example of how multiple locks
are acquired. Transaction $T_1$ requires locks on records $A$, $B$,
and $C$. As mentioned in \secref{avoidance}, to avoid deadlocks,
\systemName{} requires all transactions
to request locks from concurrency control threads in a well-defined
order.  
In \figref{multiple_locks}, this order is $CC_1$, followed
by $CC_2$, followed by $CC_3$\footnote{In \figref{multiple_locks},
each thread is assumed to have its own input queue (as in
\figref{single_lock}). However, we do not show them in the figure.}

Execution threads request the first of these concurrency control
threads to acquire locks on its partition. 
$E_1$ therefore requests $CC_1$ to acquire $T_1$'s locks (\textbf{Step
  1}). For every
lock required on later concurrency control threads, the concurrency
control thread itself forwards the request. For instance, in order to
acquire a lock on record $B$, $CC_1$ forwards $T_1$ to
$CC_2$ (\textbf{Step 2}). Similarly, after acquiring a lock on record
$B$, $CC_2$ forwards $T_1$ to $CC_3$ (\textbf{Step 3}).
The last concurrency control thread that needs to
acquire a lock on a transaction returns the transaction back to the
execution thread. In this case, $CC_3$ returns $T_1$ to $E_1$
(\textbf{Step 4}).

If this optimization were not in place, $E_1$ would have to request
locks to each concurrency control thread directly. First it would send
a message to $CC_1$ to acquire a lock on $A$. $CC_1$ then responds to
$E_1$ once it the lock on $A$ has been acquired by $T_1$. $E_1$ then
repeats this process for each concurrency control thread. The number
of messages per execution-concurrency control thread interaction is
two. Therefore, the total number of messages sent in the case of
execution thread mediated lock acquisition is $2(N_{cc})$ (where
$N_{cc}$ is the number of concurrency control threads from which the
execution thread requires locks). On the other hand, \systemName{}'s
optimized lock acquisition procedure requires $N_{cc} + 1$ messages
(one message to each concurrency control thread, and one last message
to the execution thread). The reduced number of messages is directly
correlated with decreased waiting due to asynchrony.

Note that concurrency control threads
may be subject to over- and under-utilization
due to workload skew \cite{pavlo2012skew-aware}. 
\systemName{} can re-use prior techniques for addressing utilization
imbalance in shared-nothing systems
\cite{taft2014estore,pavlo2012skew-aware,cheung2012automatic} in order to partition data
among concurrency control threads.

\subsection{Alternative architectures}
\label{sec:alternative}

In \systemName{}, the set of database objects is partitioned across
concurrency control threads. This design ensures that concurrency
control threads do not share any data, therefore avoiding data
movement and synchronization overhead (\secref{conflated}). Note,
however, that partitioning objects across concurrency control threads
is orthogonal to the design principle of separating
functionality. \systemName{}'s use of partitioning is just one
possible implementation of locking-based concurrency control. 

A plausible alternative implementation would be to share a single lock
table across all concurrency control threads. A single concurrency
control thread could then obtain all the logical locks needed by a
particular transaction. Execution threads could request any one of
several concurrency control threads to acquire locks on its behalf. 
Although such an implementation would be subject to
synchronization and data movement overhead, this synchronization is
only across the concurrency control threads --- a much smaller number
of threads than the total threads in the system. Furthermore, the
database system has 
the flexibility to limit the impact of these synchronization overheads. For
example, the system could choose to assign concurrency control threads
to execute on cores within a single NUMA socket. 

Note that this flexibility to choose the number of threads to dedicate
to each function is a direct consequence of the design principle of
separating functionality. Conventional database systems do not have
this flexibility because a single thread performs all the work entailed
in executing a transaction. Furthermore, the separation of
functionality enables a single system to support more than one
implementation of a particular sub-system (such as the partitioned
and non-partitioned lock table). Since components interact
through narrow message-passing interfaces, the actual deployed implementation
of a sub-system, such as concurrency control, can vary depending on
system parameters, such as number of cores, number of NUMA sockets,  
and so forth.


\section{Evaluation}
\label{sec:eval}
We run our experiments on a single 80-core machine, consisting of
eight 10-core Intel E7-8850 processors and 128GB of memory. The
operating system used is Ubuntu 14.04. All our experiments are
performed in memory (none of our implementations utilize secondary
storage). In all experiments, the number of threads used is equal to
the number of cores; we pin a single long running thread to each CPU
core (\secref{cc}).

We compare \systemName{} against an implementation of two-phase
locking (2PL) within the same \systemName{} transaction management
codebase.  Our 2PL implementation uses a lock-table to store
information about the locks acquired and requested by
transactions. The lock-table is implemented as a hash-table. We
implemented two important multi-core specific optimizations to improve
the scalability of our 2PL implementation. First, the lock manager
hash-table uses per-bucket latches instead of a single latch to
protect the entire table. Per-bucket latches allow our 2PL
implementation to avoid contention and overly conservative
serialization on a single global latch. In addition to per-bucket
latches, our 2PL implementation does not acquire high-level intention
locks; transactions only acquire fine-grained logical locks on
individual records. As a consequence, latch contention occurs only
when multiple threads try to acquire or release logical locks on the
same record.  Second, our 2PL implementation \textit{never} interacts
with a memory allocator. Each database thread manually manages a
pre-allocated thread-local pool of memory. Avoiding interaction with a
memory allocator (such as malloc) removes superfluous synchronization
in the operating system's memory management logic and the memory
allocator's logic. 
This allows
us to isolate the sources of synchronization overhead to those in our
own implementation.

To evaluate the overhead of deadlock handling in 2PL, we implement
three different deadlock detection/avoidance mechanisms:

\textbf{Wait-for graph.}
We use a graph to track the dependencies between transactions waiting
to acquire logical locks, and the current holders of the lock.  We
only add edges to the wait-for graph if a transaction requests a lock,
but finds that the lock is currently held in a conflicting mode by
another transaction.  The presence of a cycle in the wait-for graph
implies that the transactions that constitute the cycle have
deadlocked.  In order to scale across multiple cores, our
implementation avoids the use of a global latch to protect the entire
graph. Instead, each database thread maintains a local partition of
the wait-for graph, as is done by Yu \etal{} \cite{yu2014staring}.


\textbf{Wait die.}
Unlike the wait-for graph deadlock detection technique, which allows
transactions to deadlock, and then detects and resolves deadlocks
after the fact, wait die proactively avoids deadlocks by aborting
transactions if they are suspected to be involved in a deadlock.  In
wait die, each transaction is assigned a timestamp prior to its
execution, and the timestamp is used to determine whether or not the
transaction is allowed to wait for a logical lock.  If a transaction
fails to immediately acquire a lock, then wait die only allows the
transaction to wait on prior transactions if its timestamp is smaller
than that of the current lock holder. If not, the transaction is
aborted and restarted. Thus, wait die prioritizes older transactions
(transactions with smaller timestamps), over younger transactions
(transactions with larger timestamps).  Each database thread uses
the local timestamp counter on its CPU core to assign transactions
their timestamps. Reading from the the core-local timestamp counter is
low-overhead and contention-free. Core-local timestamp counters are
therefore a cheap scalable source of monotonically increasing timestamps. 


\textbf{Dreadlocks.}
This state-of-the-art deadlock detection technique was proposed by Koskinen \etal{} in
the context of mutual exclusion spin locks
\cite{koskinen2008dreadlocks} and is used in the multi-core optimized
version of the Shore database system (Shore-MT)
\cite{johnson2009shore}. Each transaction,
$T$, maintains a
digest, a data-structure which indicates the set of \textit{other}
transactions that $T$ waits on for locks. Intuitively, a transaction's digest is
a compact representation of its localized wait-for graph;
$T$'s digest contains the \textit{transitive closure} of the
transactions it waits for. If $T$ fails to acquire a lock, $T$
performs a set-union of its digest with the digest of the current lock
holder. If $T$ ever finds itself in its own digest, then it means that
a deadlock has occurred; since a digest corresponds to $T$'s transitive
closure in its wait-for graph, $T$ will appear in its own digest if
the transitive closure contains a cycle. Note that digests are
amenable to a simple bitmap representation, and that a particular transaction's
digest is always updated by the thread responsible for running
the transaction. However, other threads can \textit{read} the
transaction's digest. As a consequence, updates to a
transaction's digest can be performed without the use of latches
\cite{koskinen2008dreadlocks}.

We also include a version of 2PL that uses the deadlock avoidance
protocol described in \secref{avoidance}. We call this baseline
\locking{}. This deadlock-free implementation analyzes each
transaction prior to its execution in order to obtain its read- and
write-sets and acquires locks in the lexicographical order in advance
of transaction execution (as described in \secref{avoidance}).
We thus can compare our version of deadlock avoidance with three other
widely-used techniques for handling deadlock in 2PL systems.


\begin{figure} \centering
\includegraphics[width=\linewidth]{\figdir{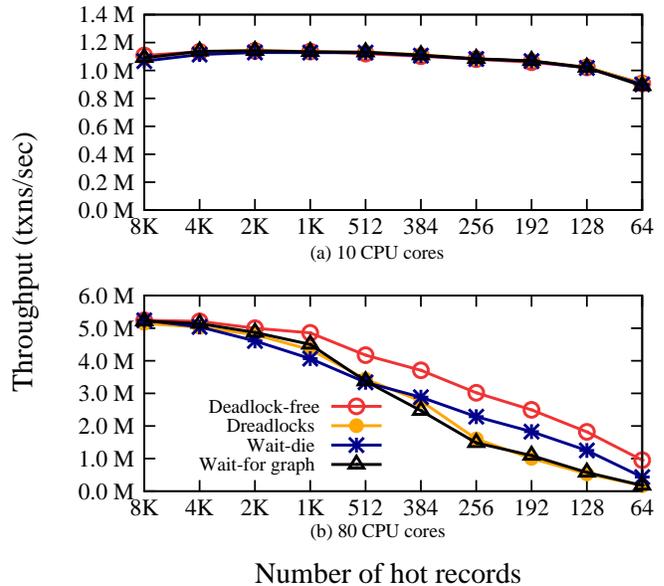}}
\subfigure{\label{fig:deadlock_10_threads}}
\subfigure{\label{fig:deadlock_80_threads}}
\vspace{-2.5em}
\caption{Throughput while varying the number of hot records in the
  database. \textbf{(a)} Number of database cores =
  10. \textbf{(b)} Number of database cores = 80.}
\label{fig:deadlocks}
\end{figure}


\subsection{Quantifying deadlock handling overhead}
\label{sec:eval_deadlocks}
We begin our evaluation with experiments to show the overhead of the
three deadlock handling mechanisms we implemented --- wait-for graph,
wait die, and dreadlocks --- and compare them with our deadlock-free
implementation of 2PL.  We measure the throughput of these
mechanisms under varying levels of contention. 

Our benchmark uses a single table of 10,000,000 records.  Each record's
size is 1,000 bytes. The workload consists of transactions that each
perform read-modify-write operations on ten records. Of the ten
records updated by each transaction, we pick two records uniformly at
random from a set of ``hot'' records, while the remaining eight are
selected uniformly at random from a set of ``cold'' records.  We vary
the level of contention in the workload by varying the number of hot
records. Intuitively, if transactions pick two records from a small
set of hot records, then the probability that they will conflict is
higher than if the set is large.  We run two experiments, the first
dedicates 10 CPU cores to the database system, while the second uses
all 80 CPU cores of our test machine.


\figref{deadlock_10_threads} shows the throughput of each deadlock
handling protocol when the database runs on 10 CPU cores.  The number
of hot records decreases as we move from left to right along the
x-axis. As a consequence, contention \textit{increases} from left to
right. Each system's throughput decreases with increasing
contention. This is expected; as contention increases, the likelihood
of conflicts between transactions increases, which in turn limits the
amount of concurrency the database system is able to exploit. However,
the relative difference in throughput between each deadlock handling
mechanism is small.

Next, we perform the same experiment with 80 CPU
cores. \figref{deadlock_80_threads} shows the results. In this case,
we find that the throughput of each system drops
\textit{significantly} with increasing contention. 

The performance of the wait-for graph mirrors that of dreadlocks across
the spectrum of contention. We attribute this to the fact that both
wait-for graph and dreadlocks effectively use the same algorithm to
detect deadlocks, they only use different data-structures to represent
the same information. As explained previously, in the dreadlocks
deadlock detection algorithm, each database thread maintains a bitmap
to represent the transitive closure of a transaction's waits-for
dependency graph. The wait-for graph deadlock detection algorithm
represents the waits-for relationship among transactions as an
explicit graph. Logically, however, the two algorithms are equivalent;
both abort transactions on detecting a cycle in their respective graph
representations. For this reason, we find that wait-for graph and
dreadlocks have nearly equivalent performance.


Wait-for graph and dreadlocks outperform wait die in the low to medium
range of contention (the left-hand-side of
\figref{deadlock_80_threads}). This is because wait die suffers from
false positives; that is, it aborts transactions despite the absence
of deadlocks.  Consider two transactions $T_0$ and $T_1$, such that
$T_0 < T_1$. If $T_0$ and $T_1$ conflict on records $x$ and $y$, and
$T_0$ has already acquired locks on both $x$ and $y$, then $T_1$'s
subsequent requests for a lock on either $x$ or $y$ will cause $T_1$
to abort and restart (because $T_1$ is younger than $T_0$). However,
because $T_0$ has already acquired locks on both $x$ and $y$, the
requests for locks on $x$ and $y$ by $T_1$ cannot result in a
deadlock. The throughput of wait die reduces because of the wasted
work due to such unnecessary aborts. However, under low contention, 
the rate of conflicts among transactions is low. Thus, 
although wait die is outperformed by wait-for graph and dreadlocks,
the difference is modest.


While wait-for graph and dreadlocks outperform wait die under low
contention, this trend
reverses under high contention. We attribute this to the fact
that deadlock detection in wait die is less expensive than that in
wait-for graph and dreadlocks. Wait die's deadlock handling involves
comparing the timestamps of two conflicting transactions, while
wait-for graph and dreadlocks maintain additional data-structures for
the purposes of deadlock detection. Furthermore, wait die aborts
transactions as soon as it notices a violation of its timestamp
rule. In contrast, wait-for graph and dreadlocks wait until they
notice a cycle in the waits-for graph of transactions. The result
is that wait die aborts doomed transactions earlier than the wait-for
graph and dreadlocks deadlock detection techniques.

Most importantly, \figref{deadlock_80_threads} indicates that
\locking{} \textit{always} outperforms all three deadlock
handling mechanisms. There are two reasons for this. First, when
deadlocks do not occur too often (in the middle of the graph),
deadlock-free locking outperforms the other schemes due to its
low-overhead. \locking{} only has to analyze transactions'
read- and write-sets in advance, and request locks in the correct
order, while the other schemes are burdened by running deadlock
handling logic (wait-for graph and dreadlocks) and aborting
transactions due to false positives (wait die). Second, since
deadlocks themselves do not occur, \locking{} does not suffer from aborts, and
subsequent wasted work and overhead due to retries.  
\locking{}'s advantage over deadlock handling techniques
grows with increasing contention. At the right-most point in the
graph, deadlock-free locking's throughput is 2.2x, 5.5x and 5.5x times
that of wait die, dreadlocks and the wait-for graph, respectively.



\figref{deadlock_80_threads} validates the fact that under high
contention workloads, deadlock handling logic is a significant
impediment to multi-core throughput; dynamic deadlock handling techniques
are \textit{always} outperformed by \locking{}. Furthermore, the fact that
throughput drops so drastically from 10 to 80 CPU cores indicates the
problem will only get worse with increasing core counts.

\subsection{Tradeoffs in thread allocation}
\label{sec:conc_scalability}

\begin{figure} 
\centering
\includegraphics[width=\linewidth]{\figdir{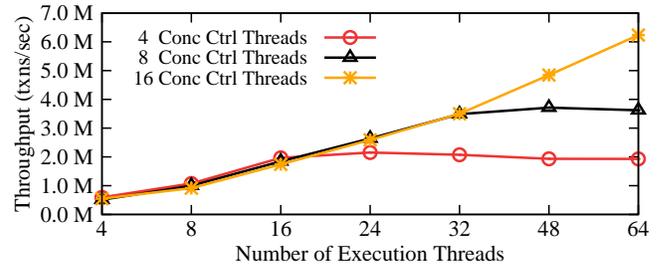}}
\caption{Execution thread scalability in \systemName{} under various fixed concurrency
  control thread configurations.}
\label{fig:vary_cc}
\end{figure}
\systemName{}'s partitioned functionality means that each database
thread can be assigned one of two roles; concurrency control or
transaction execution. Given a fixed number of threads, \systemName{}
must apportion threads to either concurrency control or execution.
This section shows the performance implications of various concurrency
control and execution thread allocations. We run one experiment, in
which 80 threads (corresponding to our test machine's 80 physical CPU
cores) are made available to \systemName{}.  We experiment with
multiple \systemName{} configurations. In each configuration, we fix
the number of concurrency control threads, and measure throughput
while varying the number execution threads.

We configure the database with a single table of 10,000,000 rows, each
of size 1,000 bytes. The workload consists of transactions performing
10 read-modify-write operations on unique records. The records in
transactions' read- and write-sets are selected uniformly at random
from the set of 10,000,000 database rows. Each transaction acquires
all its locks from a single concurrency control thread (we experiment
with transactions that acquire locks on multiple concurrency control
threads in \secref{partitioned_eval}, and \secref{tpcc}).

\figref{vary_cc} shows the results of the experiment. We experiment
with three concurrency control configurations, each corresponding to a
curve in \figref{vary_cc}. In each concurrency control configuration,
throughput initially increases with increasing execution thread
count. This is because on the left-hand-side of \figref{vary_cc},
there are not enough execution threads to fully utilize the available
concurrency control threads. Throughput continues increasing until a
sufficient number execution threads saturate the available concurrency
control threads. At this point, throughput plateaus. The point at
which each curve plateaus is directly proportional to the number of
concurrency control threads; more concurrency control threads can
sustain a higher aggregate throughput than fewer concurrency control
threads.

While \systemName{} provides the flexibility to configure the number
of concurrency control and execution threads, the choice of the
optimal division of threads between concurrency control and execution
is not obvious. Too few execution threads causes under-utilization of
concurrency control threads, and vice-versa. Fortunately,
\systemName{} uses a staged event driven architecture (SEDA)
\cite{welsh2001seda}; \systemName{}'s concurrency control and
execution modules correspond to SEDA stages communicating via
explicit-message passing. Systems based on SEDA are amenable to
dynamic allocation of resources (such as threads) based on load. In
order to decide on the optimal allocation of threads between
concurrency control and execution, therefore, \systemName{} can use
techniques for dynamic resource allocation on SEDA systems.

\subsection{Multi-partition transactions}
\label{sec:partitioned_eval}
This section explores the cost of multi-partition transactions
(transactions that need locks located in multiple concurrency control
threads) in \systemName{}. We compare performance against \locking{}
(which has no partitioning whatsoever), and a fully-partitioned,
``shared-nothing'' system, where no memory is shared between
partitions. Our implementation is based on the single-node
(not distributed) version of the architecture of
H-Store/VoltDB and HyPer
\cite{hstore,kemper11hyper}. This \partitionedStore{} baseline is similar to the 
corresponding implementation by Tu \etal{} in Silo
\cite{tu2013speedy}. \partitionedStore{} physically partitions data
across database worker threads, such that each worker has its own
local hash-table index. For concurrency control, \partitionedStore{}
associates a coarse-grain partition-level spinlock with each
worker. In order to execute a transaction, a worker thread obtains
partition-level spinlocks on every partition that the transaction
needs to access. If every transaction is single-partition, the
corresponding workers will only acquire their own partition-level
spinlocks. In the single-partition case, therefore, workers' lock
acquisitions never conflict with each other. Furthermore, in the
single-partition case, 
spinlock acquisition has minimal overhead because the lock is cached
by the corresponding worker thread.

We run two experiments, both highlighting
the effect of multi-partition transactions on performance. The first
experiment compares the throughput of each system while varying the
number of partitions accessed by each transaction. The second varies
the fraction of multi-partition transactions in the workload.

The experiments in this section use a database which consists of a
single table with 10,000,000 records of size 1,000 bytes each. These
10,000,000 records are uniformly spread across \partitionedStore{}'s
physical partitions. Similarly, 10,000,000 logical locks are
uniformly spread across \systemName{}'s concurrency control threads.
Both experiments use transactions which perform 10 read-modify-write
operations. In \partitionedStore{}, multi-partition transactions span
physical partitions. Similarly, multi-partition transactions in
\systemName{} request locks from multiple concurrency control
threads. For instance, a transaction which accesses three physical partitions
in \partitionedStore{} will request locks from three concurrency
control threads in \systemName{}.

\begin{figure} \centering
\includegraphics[width=\linewidth]{\figdir{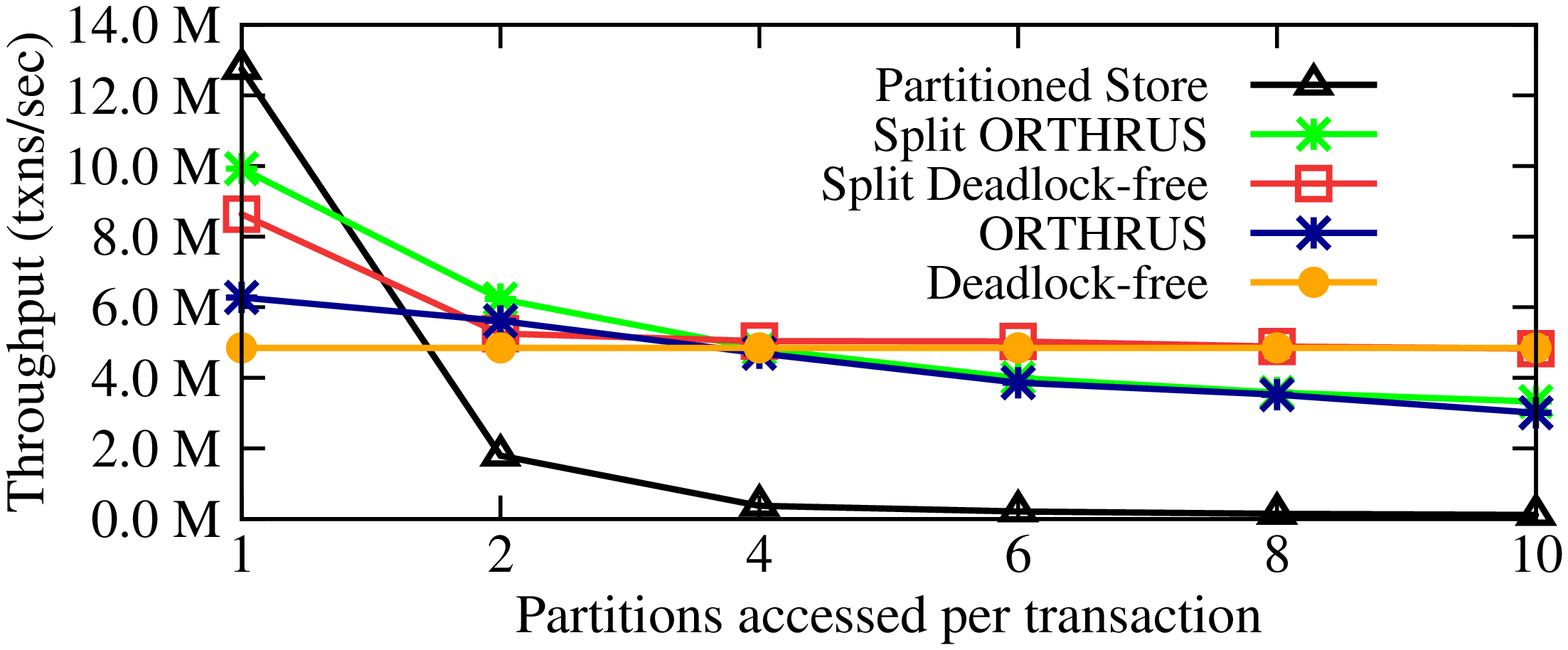}}
\caption{Performance of \systemName{} and \partitionedStore{} as
the number of partitions accessed per transaction is varied.}
\label{fig:vary_partitions}
\end{figure}

\textbf{Vary partitions per transaction.} 
\figref{vary_partitions} shows the performance of
\systemName{}, \partitionedStore{} and \locking{} while
varying the number of partitions accessed by each transaction. When transactions
are restricted to a single partition,
\partitionedStore{} outperforms \systemName{} and \locking{}. 
There are two primary reasons for this. First, \partitionedStore{} requires no concurrency control
when transactions are restricted to a single
partition. Second, \partitionedStore{} index structures have better
cache locality because indexes are physically partitioned across
worker threads.  As transactions access two or more partitions, however
\partitionedStore{} experiences a sharp drop in
throughput. \partitionedStore{}'s drop in throughput is due to its use
of coarse-grain concurrency control. \partitionedStore{}
isolates transactions at the level of partitions; a pair of
transactions conflict if they both access the same partition. In
contrast, \systemName{} and \locking{} isolate transactions
at the granularity of read- and write-conflicts on individual records.

We also find that \systemName{}'s throughput decreases as transactions
access more partitions. However, unlike \partitionedStore{},
\systemName{}'s drop in throughput is more modest. \systemName{}'s
throughput drops because of the increase in the number of messages
required to acquire a transaction's locks. In the single-partition
case, each transaction acquires its locks from a single concurrency
control thread. However, as transactions' locks are distributed over
a greater number of concurrency control threads, the number of message
hops during lock acquisition increases. In particular, the number of
messages required to acquire a transaction's locks is equal to
$N_{cc}+1$, where $N_{cc}$ is the number of concurrency control
threads on which a transaction's locks reside
(\secref{optimizations}). Clearly, as $N_{cc}$ increases, the number
of messages involved in acquiring a single transaction's locks
increases. \figref{vary_partitions} also shows that \locking{}'s
throughput remains unchanged as the number of partitions
accessed by a transaction increases. Deadlock-free locking is a
shared-everything system, and hence is not subject to additional overhead in the
presence of multi-partition transactions.

To better understand the performance characteristics of the curves in
\figref{vary_partitions}, we physically partitioned indexes across
\systemName{} and \locking{}'s worker threads\footnote{Tu \etal{}
perform a similar analysis in Silo \cite{tu2013speedy}}. These curves
are marked \splitSystemName{} and \splitLocking{},
respectively.  This optimization puts all three systems on the
same level as
\partitionedStore{} with respect to cache locality,  and any remaining
difference between \partitionedStore{} and these two
new curves can be attributed to concurrency control. 

When analyzing the difference between \partitionedStore{} and
\splitSystemName{}/\splitLocking{} and comparing this difference to
the difference between \partitionedStore{} and
\systemName{}/\locking{}, we can conclude that when transactions
are restricted to
a single partition, \partitionedStore{}'s main advantage
over \systemName{} and \locking{} is due to the smaller cache
footprint of partitioned indexes.
\partitionedStore{}'s throughput is about 2x and 2.6x that of
\systemName{} and \locking{}, respectively, while its advantage over
\splitSystemName{} and \splitLocking{} is a more modest 1.3x
and 1.5x, respectively.  Furthermore, as transactions access more
partitions, the performance of the partitioned variants of
\systemName{} and \locking{} converge to their non-partitioned
counterparts. This confirms that \partitionedStore{}'s poor
performance under multi-partition transactions is due to its use of
coarse-grain concurrency control.

It should be noted that a big
advantage of \systemName{} and \locking{} over \partitionedStore{} is
that these systems by default do not require a user to be concerned
about finding a near-perfect data partitioning such that the vast
majority of a transactions in a workload will only access a
single-partition.  However, if a good or near-perfect partitioning is
available for a
particular workload, there is no
reason why \systemName{} and
\locking{} cannot benefit from it by partitioning their indexes
across worker threads accordingly. In other words, although our
primary motivation for introducing \splitSystemName{} and
\splitLocking{} was in order to break down the performance differences
between \partitionedStore{} and \systemName{}/\locking{}, if a good
data
partitioning is available for a workload, \splitSystemName{} could be
used instead of \systemName{}, and achieve much closer performance to
partitioned stores on single-partitioned transactions, while maintaining
its significant advantages over partitioned-stores for multi-partition
transactions.

\textbf{Vary fraction of multi-partition transactions.}
Although \figref{vary_partitions} shows the effect of multi-partition
transactions on each system's throughput, realistic workloads involve
a mix of single-partition and multi-partition transactions. In this
experiment, we evaluate a workload consisting of both single- and
multi-partition transactions. We vary the percentage of
multi-partition transactions in the workload. Multi-partition
transactions run on exactly two
partitions. 

\begin{figure} \centering
\includegraphics[width=\linewidth]{\figdir{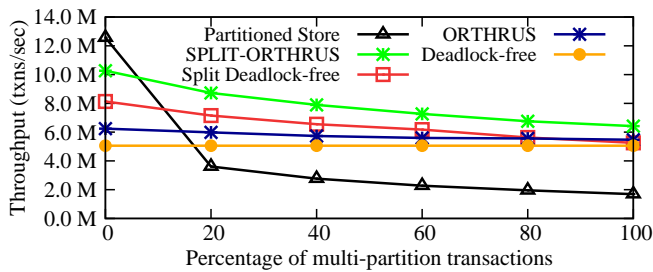}}
\caption{Performance of \systemName{} and \partitionedStore{} as
the percentage of multi-partition transactions is varied.}
\label{fig:vary_partitions_pct}
\end{figure}

\figref{vary_partitions_pct} shows the result of the experiment.  As
in the previous experiment, we find that \partitionedStore{}
outperforms \systemName{} and \locking{} when all transactions are
single-partition (0\% multi-partition
transactions). \partitionedStore{}'s throughput decreases as the
fraction of multi-partition transactions increases. This is expected
because we already saw in \figref{vary_partitions} that
\partitionedStore{}'s performance decreases dramatically as soon an
transactions access more than one partition. 

\systemName{}'s throughput also
decreases as the fraction of multi-partition transactions
increases. As elaborated in the previous experiment, this is 
due to the increase in the number of messages required to acquire
transactions' locks. However, despite the
decrease in throughput, \systemName{} \textit{always} outperforms
\locking{} (even when the percentage of multi-partition transaction is
100\%). Although both \systemName{} and \locking{} use the same
underlying locking-based protocol, \systemName{}'s partitioned
functionality allows concurrency control and execution threads to
better utilize data- and instruction-caches (\secref{cc}).

Finally, \figref{vary_partitions_pct} also evaluates
\splitSystemName{} and \splitLocking{}. These curves reinforce the
fact that the decrease in concurrency in \partitionedStore{} far
outweighs the reduced cache-locality due to transactions accessing
multiple physical partitions.

\subsection{Performance under contention}
\label{sec:tpcc}
This section evaluates the performance of \systemName{} under
contention. We compare \systemName{} against \locking{}, and 2PL with
dreadlocks. The experiments in this section run a subset of the TPC-C
benchmark. As in common practice, our TPC-C implementation does not
model client ``think'' time, transactions are executed as one-shot
stored procedures
\cite{tu2013speedy,bailis2014avoidance,ren13lightweight,hstore}. We
restrict our evaluation to TPC-C's NewOrder and Payment
transactions. These two transactions make up the vast majority of the
benchmark; approximately 45\% and 43\% of transactions in the full
TPC-C mix are NewOrder and Payment transactions, respectively.
Furthermore, NewOrder and Payment are short update
transactions\footnote{Compared to long running transactions such as StockLevel}, and
thereby put greater stress on concurrency control.  Our evaluation
therefore uses an equal mix of NewOrder and Payment transactions; both
types of transaction are equally likely to occur.

TPC-C's database conforms to a tree-based schema \cite{hstore}. Most
tables in TPC-C have a foreign key dependency on the ``root''
Warehouse table. Excluding the Warehouse table itself, out of eight
tables in the TPC-C database, only one, Item, does not contain a
foreign key dependency on the Warehouse table.  TPC-C's Item table is
read-only, hence, none of our baselines perform any concurrency
control on reads to Item table's rows.  In TPC-C, the $warehouse\_id$
attribute is the primary key of the Warehouse table, and foreign key
in all other tables (apart from Item).  \systemName{} partitions
database tables across concurrency control threads based on each row's
$warehouse\_id$ attribute.  Note that our evaluation adheres to the
TPC-C specification of requiring 10\% of NewOrder and 15\% of Payment
transactions to span two warehouses. Therefore, approximately 12.5\%
of transactions in our evaluation will require locks from two
concurrency control threads. 
Some Payment transactions' read- and write-sets are deducible only
upon reading the value of a secondary index. In particular, 60\% of
Payment transactions must find a Customer by a secondary index on
customers' last name. For this subset of Payment transactions,
\systemName{} must speculatively read this secondary index in order to
obtain the transaction's read- and write-sets using the OLLP protocol
described in \secref{avoidance}.


\subsubsection{Throughput under varying contention}
\begin{figure}
\centering
\includegraphics[width=\linewidth]{\figdir{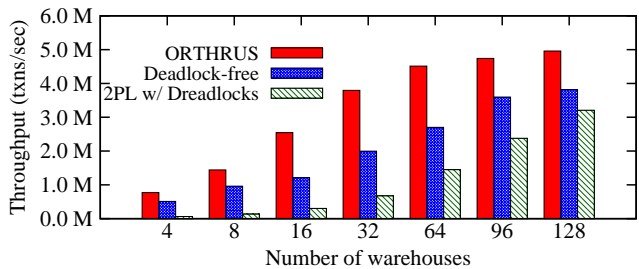}}
\caption{TPC-C NewOrder and Payment throughput, varying number of
  warehouses. Each system uses 80 CPU cores.} 
\label{fig:tpcc_vary_wh}
\end{figure}

Since the TPC-C database schema is tree-based, and rooted at the
Warehouse table, contention on all tables (except the read-only Items
table) can be controlled by varying the number of records in the
Warehouse table; decreasing the number of warehouses increases the
level of contention in the workload. \figref{tpcc_vary_wh} shows the
throughput of each system while varying the number of warehouses. The
number of warehouses increases from left to right along the x-axis,
therefore, contention \textit{decreases} from left to right.

When the number of warehouses is small, we find that \systemName{}
significantly outperforms 2PL. When the
number of warehouses is small, both transactions update highly contended records.
Payment updates two records, one each from the Warehouse and District
tables. NewOrder updates a single District record. Deadlock free
locking suffers from latching overhead associated with acquiring
logical locks on each of these records (synchronization
overhead). Furthermore, the linked-list corresponding to a 
bucket in the lock table must also be moved across cores (data
movement overhead).

In addition to suffering from the same sources of overhead as
deadlock-free locking, 2PL with dreadlocks must also execute deadlock
handling logic,
which further reduces throughput. Dreadlocks requires all transactions
waiting on a particular logical 
lock to spin on the lock holder's digest. When the lock holder
eventually releases the lock, it updates its digest, which propagates
to the threads spinning on the digest. In order to read the new value
of the digest, the readers' cached values of the digest must be
invalidated and then reloaded with the new value. Note that this cache
coherence overhead is \textit{in addition} to the cache coherence
overhead associated with latch acquisition and linked-list traversal
experienced by deadlock free locking. Furthermore, we found that the
actual occurrence of deadlocks on the TPC-C workload was rare. 2PL with
Dreadlocks is
thus subject to severe cache coherence overhead despite the fact that
the workload itself is deadlock free.

Unlike 2PL, \systemName{} does not
experience cache-coherence induced overhead. In \systemName{}, a
single concurrency control thread is responsible for processing every
lock operation on a particular record. Concurrency control threads
therefore require no synchronization to process lock acquisition or
release requests. Furthermore, the meta-data associated with the
locks (the linked list of lock requests) does
not need to move across cores. On the contrary, since Warehouse and
District records are popular, the linked lists of lock requests on
these records experience good cache locality on concurrency control
threads. As mentioned previously,  
10\% of NewOrder and 15\% of Payment transactions to span
two warehouses. Since we assign a single concurrency control thread to
a particular warehouse, this subset of NewOrder and Payment
transactions must interact with two concurrency control
threads. These transactions are therefore subject to greater lock hold
times on popular records due to asynchrony (\secref{optimizations}).
Despite the longer duration for which contended locks are held,
\systemName{} is able to significantly outperform both locking implementations.

As the number of warehouses increases, the level of contention in the
workload decreases. As a consequence, both locking systems experiences
lower synchronization overhead.  However, despite the decrease in
contention, we find that \systemName{} maintains a significant
advantage over both locking systems.  At 128 warehouses,
\systemName{}'s throughput is 1.3x and 1.5x that of \locking{} and
2PL, respectively.  We attribute \systemName{}'s advantage to the
lower instruction- and data-cache footprint entailed by partitioned
functionality (\secref{conflated} and \secref{cc}).

\subsubsection{Scalability under high contention}
\begin{figure} \centering
\includegraphics[width=\linewidth]{\figdir{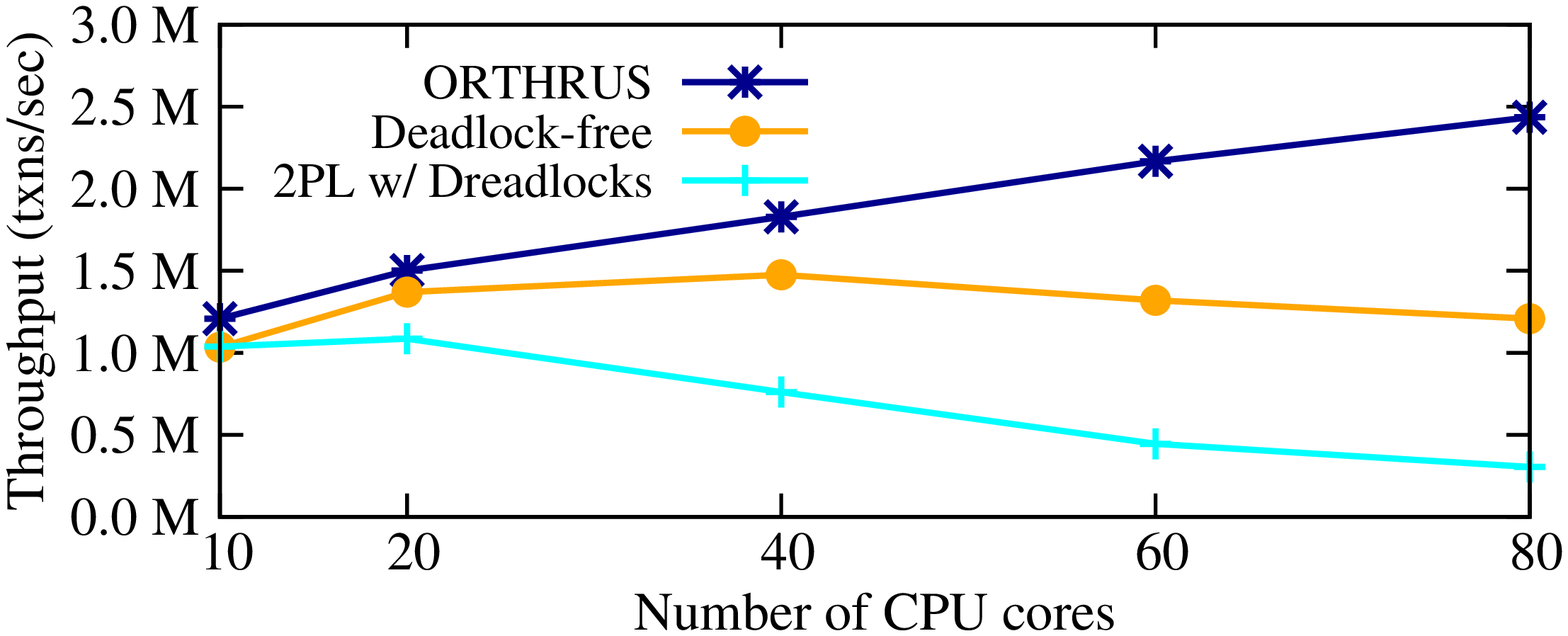}}
\caption{TPC-C throughput while increasing core count. Number of warehouses = 16.}
\label{fig:tpcc_vary_cores}
\end{figure}

Next, we compare the scalability of
\systemName{}, \locking{} and 2PL under high contention. We set the
total number of warehouses to 16.  \figref{tpcc_vary_cores} shows the
results of the experiment. When each system uses 10 CPU cores, we find
that \locking{} and 2PL's throughput is identical, validating that
there are no implementation specific bottlenecks in our 2PL baseline.
On increasing the number of CPU cores, we find that 2PL's throughput
begins to drop because of the overhead of deadlock handling logic. As
mentioned previously, deadlocks occur with negligible frequency on the
TPC-C benchmark. However, despite the negligible frequency of
deadlocks, we find a significant difference between dreadlocks and
deadlock-free locking; this difference validates our claim that
deadlock handling logic is a significant source of overhead,
even on workloads which are devoid of deadlocks.  Despite the
significant level contention in the workload \systemName{}'s
throughput scales with additional CPU cores. At 80 cores,
\systemName{} outperforms \locking{} and 2PL by 2x and nearly an
\textit{order of magnitude}, respectively.

\subsubsection{Execution time breakdown}
\label{sec:tpcc_breakdown}
\begin{figure} \centering
\includegraphics[width=\linewidth]{\figdir{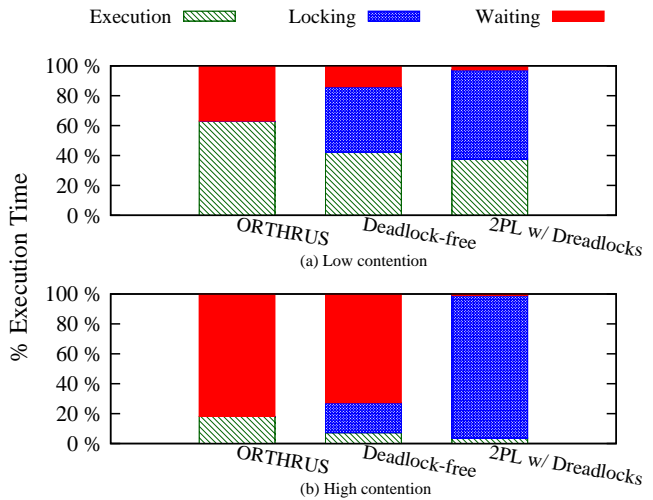}}
\subfigure{\label{fig:128_breakdown}}
\subfigure{\label{fig:16_breakdown}}
\vspace{-2.5em}
\caption{Execution thread CPU time breakdown on TPC-C with 80 threads. \textbf{(a)}: 128 Warehouses (low contention). \textbf{(b)}: 16 Warehouses (high contention)}
\label{fig:breakdown}
\end{figure}

We conclude our experimental evaluation by showing the breakdown of
CPU time on database execution threads in \systemName{}, \locking{},
and 2PL. \figref{breakdown} shows that 2PL spends significantly more
time locking records than \locking{}. This is due to
the fact the the dreadlocks algorithm spends time spinning on
threads' digests in the lock manager. The difference between these
two baselines is only that 2PL spends time waiting within the lock
manager, while \locking{} spends time waiting outside the lock
manager. 

\systemName{}'s execution
threads spend more cycles doing useful work than both \locking{} and
2PL under both high contention and low contention. Under high
contention all three systems spend a large fraction of their time
waiting on locks. However, \systemName{}'s execution threads spend
significantly more time executing transactions (in comparison to
\locking{} and 2PL).  In \figref{16_breakdown}, \systemName{}'s
execution threads spend 18\% of their time executing transactions. In
comparison, \locking{} and 2PL threads spend just 7.2\% and 3.7\% of
their time doing useful work. \systemName{}'s execution thread
utilization is about 2.5x and 5x greater than that \locking{} and 2PL,
respectively. It should be noted that as a consequence of its
partitioned functionality, \systemName{} uses fewer execution threads
than \locking{} and 2PL. In both experiments, \systemName{} uses 16
concurrency control threads and 64 execution threads. In contrast, the
other systems utilize all 80 threads for execution. However, despite
having 20\% fewer execution
threads at its disposal, \systemName{} outperforms \locking{} and 2PL
by much better utilizing its available execution threads.

\section{Related Work}
\label{sec:related}
\textbf{Synchronization.}
Remote Core Locking (RCL) is a technique for reducing synchronization
overhead and increasing data locality in contended critical sections
\cite{lozi2012remote}. RCL classifies a subset of a machine's CPU
cores as ``server'' cores, and partitions contended critical sections
across these server cores. Threads request server cores to execute
critical sections on their behalf.  Flat combining is a
synchronization technique that addresses the same problem as RCL; the
impact of contended critical sections on synchronization overhead and
data locality \cite{hendler2010flat}. Unlike RCL, flat combining does
not dedicate CPU cores for the sole purpose of critical section
execution. Instead, a single ``combiner'' thread is dynamically chosen
to execute critical sections on behalf of others. Both RCL and flat
combining use the design principle of dedicating a single thread to
repeatedly execute critical sections for the purposes of exploiting
instruction- and data-cache locality, and reducing synchronization
overhead.  \systemName{} uses the same design principle; RCL's server
cores, and flat combining's combiner are analogous to \systemName{}'s
concurrency control threads. However, both RCL and flat combining
address high contention in critical sections without regard to any
higher level functionality. In contrast, \systemName{}'s concurrency
control threads implement logical locking, and must therefore
carefully address harmful interactions between logical locking and
asynchronous message-passing (\secref{optimizations}).


\textbf{Operating systems.} Baumann \etal{} design a
multi-core operating system kernel, Barrelfish, that forbids
shared-memory-based inter-core communication altogether
\cite{baumann2009multikernel}. Instead, Barrelfish forces cores to
communicate using explicit message passing. One of the primary
motivations for their work was the inability to reason about
contention in conventional shared-memory operating system kernels.
Wentzlaff \etal{} propose a factored operating system (fos), which
partitions operating system functions across the cores of a single
machine. fos's design is intended to reduce contention, and improve
instruction- and data-cache locality in operating system kernels
\cite{wentzlaff2009factored}. \systemName{} shares some of the same
motivation as Barrelfish and fos, namely, addressing 
conflated functionality overhead (\secref{conflated}).  However, Barrelfish and
fos address overheads in operating system kernels, while
\systemName{}'s design addresses overheads in database concurrency
control.

\textbf{Partitioned functionality.}
Bernstein and Das, and Ding \etal{} propose distributed DBMSs in which OCC validation is
performed by a dedicated set of processes, independent from database
workers \cite{ding2015centiman,bernstein2015scaling}.
Faleiro \etal{} propose techniques for lazy transaction evaluation
\cite{faleiro14lazy} and multi-version concurrency control
\cite{faleiro2015multiversion}. Like \systemName{}, the design of 
these systems involved separating concurrency control and transaction
execution modules. 
However, none of these systems
employed explicit message-passing as an inter-thread communication
technique on a single multi-core machine. 
Moreover, this paper analyzes the broader implications of
separating concurrency control from transaction execution, while these
prior systems used separate concurrency control and execution threads
in narrower context of distributed OCC, lazy transaction evaluation, and
multi-version concurrency control. 

\textbf{Staged DBMSs.}
Harizopoulos \etal{} identify several sources of overhead with
thread-based query execution architectures \cite{stageddb}. Chief
among these was poor instruction- and data-cache locality due to
thread context switching. To address this overhead, they proposed
StagedDB, a staged event-driven query execution architecture. StagedDB
uses long-lived threads to exploit inter-query instruction- and
data-cache locality.  \systemName{} uses a similar design; it
dedicates long-lived threads to perform concurrency control and
transaction execution logic respectively. These threads are pinned to
physical CPU cores in order to exploit instruction- and data-cache
locality.  \systemName{} differs from StagedDB in that it addresses
bottlenecks in transaction processing, not query
execution. Furthermore, \systemName{} addresses transaction processing
overheads on modern multi-core machines, such as synchronization, data
movement, and deadlock handling. 

Several distributed DBMSs have adopted a staged event driven
architecture (SEDA)
\cite{decandia2007dynamo,huebsch2005pier,lakshman2010cassandra,yuan2014rubato}.
SEDA is a natural fit for distributed database systems, which must
\textit{necessarily} use message-passing as a communication mechanism. 
In contrast, \systemName{} uses a staged message-passing among threads
on a \textit{single node}. 

\textbf{Multi-core optimized DBMSs.}
Several researchers have recently proposed techniques to address
synchronization overhead and improve cache-locality in multi-core
databases
\cite{lomet2013bw,wang2014transactional,neumann2015multiversion,leis2014exploiting,kimura2015foedus,porobic2012islands,tozun2014addict,levandoski2015deuteronomy}.
Johnson \etal{} devised a technique to reduce the frequency of
contended latch acquisitions in conventional lock managers
\cite{johnson2009improving}. Their technique, speculative lock
inheritance, passes contended logical locks between transactions
without requiring calls to the lock manager (consequently decreasing
the frequency of contended latch acquisition within the lock
manager). Jung \etal{} devised scalable latch-free algorithms for
conventional lock managers \cite{jung2013scalable}.  Larson \etal{}
address several scalability bottlenecks in both pessimistic and
optimistic concurrency control protocols
\cite{larson2011concurrency}. For instance, their optimistic
validation protocol does not require the use of a global critical
section (as required by conventional protocols
\cite{kung81optimistic}). Tu \etal{} propose Silo, an optimistic
main-memory multi-core database system designed to eliminate contended
centralized data-structures \cite{tu2013speedy}. All of these systems
use a conventional shared-memory design. In contrast, \systemName{}
mediates communication among concurrency control and execution threads
using explicit message-passing.

\textbf{Semantics-aware concurrency control.}
Several researchers have argued for reasoning about conflicts using
the semantics of operations
\cite{bailis2014avoidance,clements2013commutativity,conway2012lattices,roy2015homeostasis,oneil1986escrow}. Doppel
is a main-memory database system that exploits commutative operations
on highly contended records \cite{narula2014phase}. Highly contended
records are replicated across a machine's cores, and commutative
operations are satisfied by any core. In order to process
non-commutative operations, the updates on a replica are periodically
aggregated. For operations that do not commute, however, Doppel uses
conventional concurrency control. Unlike Doppel, \systemName{} is
designed to address high contention workloads in which commutativity
cannot be exploited.

\textbf{Shared-nothing systems.}
Prior research found that the cost of two-phase locking was
prohibitively expensive on main-memory database systems. Several
researchers subsequently recommended doing away with two-phase locking
altogether, and advocated for serial transaction execution. The
H-Store is an example of a database system which employs serial
transaction execution, and works best when workloads are perfectly
partitionable \cite{hstore}. HyPer is another example of a
single-thread transaction processing system, but its design is meant
to simultaneously support OLTP and OLAP workloads
\cite{kemper11hyper}. Since these systems do away with concurrency
control altogether, they do not suffer from any of the overheads
described in this paper. However, for the same reason, they cannot
adequately utilize multi-core systems when a workload contains a
non-trivial fraction of distributed transactions. 

Pandis \etal{} propose a shared-nothing transaction processing
architecture to avoid contention on locking meta-data (DORA)
\cite{pandis2010data} and indexes (PLP) \cite{pandis2011plp}.  Unlike
conventional designs where a single thread performs a transaction's
logic, their work proposes that a transaction is collectively
processed by the partitions on which it must execute. A transaction's
logic is broken into smaller sub-transactions such that an entire
sub-transaction is restricted to a single partition. Unlike
\systemName{}, DORA assigns a single thread to perform both
concurrency control and execution within a partition. Furthermore,
this paper advocates for \textit{partitioned functionality}, which
does not necessarily preclude shared-data among concurrency control or
execution threads (\secref{alternative}).  PLP is complimentary to our
work; we do not address index contention.

\section{Conclusions}
The vast majority of database systems adhere to the design principle
of assigning a single thread the responsibility of performing all logic 
on behalf of a transaction. This design principle leads to severe
scalability problems on main-memory multi-core databases due to
synchronization overhead, data movement overhead, and cache
pollution. Furthermore, these systems allow dynamic access of data
which necessitates expensive deadlock handling
mechanisms. \systemName{} addresses these limitations 
by partitioning functionality across
a machine's cores and eliminating the need for handling deadlocks. Our
experimental evaluation shows that \systemName{}'s design enables it
to outperform conventional database systems by upto an order of
magnitude on high contention workloads. 
\\ \\
\noindent {\bf Acknowledgements.} This work was sponsored by the NSF
under grant IIS-1527118. We thank the anonymous SIGMOD 2016 reviewers
for their insightful feedback.

\bibliographystyle{abbrv}
\bibliography{ref}

\begin{appendix}

\section{YCSB evaluation}
\label{sec:ycsb_eval}
In this set of experiments, we compare \systemName{}'s throughput
against that of \locking{} and 2PL with dreadlocks on the
Yahoo! Cloud Serving Benchmark (YCSB) \cite{cooper10ycsb}. 

This set of experiments uses a single table of 10,000,000
records, each 1,000 bytes in size. 
Since YCSB executes transactions over a single table, it implicitly
assumes a flat schema. However, a significant subset of OLTP workloads
instead adhere to a \textit{tree schema}, where the database consists
of a single \textit{root} table, and most tables have a foreign-key
dependency on the root table \cite{hstore}. The key difference between
a flat schema and a tree schema is that it is easy to partition data
in a tree schema such that multi-partition transactions are rare.

To model both types of schemas,
we run \systemName{} under three different configurations; single
partition, dual partition, and random. In the single partition
configuration all the locks required by a particular transaction are
guaranteed to reside on a single concurrency control thread. The
single partition configuration represents a perfectly partitionable
tree schema. In the dual partition configuration, all the locks
required by a particular transaction are guaranteed to reside on
exactly two concurrency control threads. The dual partition
configuration forces \textit{every} transaction to acquire locks from
exactly two concurrency control threads. The dual partition
configuration thus represents a workload in which \textit{every} transaction is
distributed. Under the random configuration, a transaction's locks are
randomly distributed across concurrency control threads and could
potentially access many more than two threads.

We compare \systemName{}'s throughput against that of \locking{} and
2PL with wait-die\footnote{We also evaluated the throughput of 2PL
with the wait-for graph, and dreadlocks algorithms, but found that
wait-die outperformed both these schemes.}.  Our evaluation uses two
types of transactions. The first type of transaction is read-only, and
performs 10 reads.  The second type of transaction performs 10
read-modify-write (RMW) operations.  The read-only transactions
highlight the overhead of conflated functionality on conventional
concurrency control protocols. 
The 10RMW transactions highlight the overhead of the
combination of conflated functionality and deadlock handling logic.

\subsection{Read-only transactions}
\label{sec:ycsb_readonly}
This experiment compares \systemName{}'s throughput against that of
2PL and \locking{} on a workload consisting of read-only
transactions. In this experiment, each transaction reads 10
records. We run this experiment under low and high contention. In the
low contention experiment, all 10 records in a transaction's read-set
are selected uniformly at random from the set of 10,000,000 records.
In the high contention experiment, each transaction picks 
two records from a set of hot records, and the remaining eight records
from a set of cold records (locks on two hot records are acquired
before locks on cold records). We set the number of hot records to 64.
In both low, and high contention experiments, 2PL's deadlock handling logic is never
invoked. This is because read-only transactions do not conflict with
each other. 

\begin{figure} \centering
\includegraphics[width=\linewidth]{\figdir{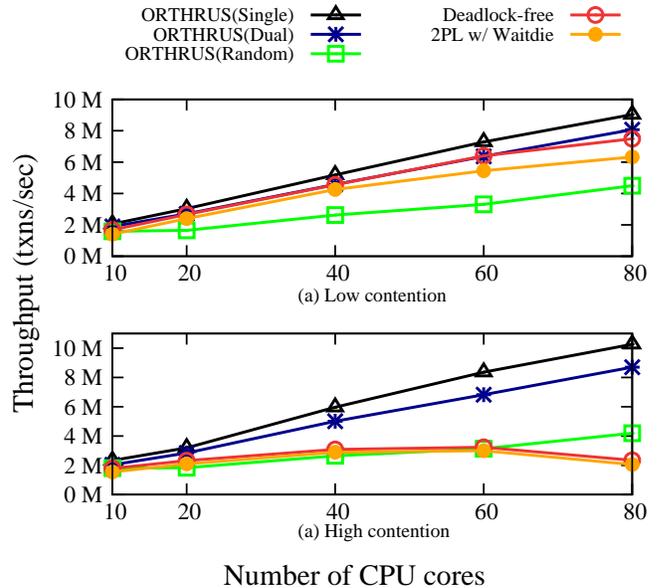}}
\subfigure{\label{fig:ycsb_readonly_low}}
\subfigure{\label{fig:ycsb_readonly_high}}
\vspace{-2em}
\caption{YCSB read-only transaction scalability. \textbf{(a)} Low
contention. \textbf{(b)} High contention.}
\label{fig:ycsb_readonly}
\end{figure}

\figref{ycsb_readonly_low} shows the results of
the low contention experiment. Read-only transactions do not write
database records, and are therefore extremely short. As a consequence,
concurrency control overhead takes up a non-trivial fraction of each
transaction's total execution time.  Single partition \systemName{}
outperforms all other baselines, which indicates that its concurrency
control overhead is the least among all the other baselines when only
one concurrency control thread is involved in a transaction. This is because
it doesn't require synchronized access to shared memory concurrency
control meta-data.

\secref{optimizations} explained that the number of message delays
required in order for a transaction to acquire its locks in
\systemName{} is $N_{cc}+1$, where $N_{cc}$ is the number of
concurrency control threads on which a transaction's locks reside. The
number of message delays required for lock acquisition directly
affects the time spent acquiring locks; more message delays correspond
to more time spent acquiring locks.  Since each transaction in single
partition \systemName{} acquires \textit{all} its locks from a single
concurrency control thread, the number of message delays per
transaction in single partition \systemName{} is 2. The
number of message delays per transaction in dual partition
\systemName{} is 3 (because every transaction must acquire its locks
from two concurrency control threads, but the threads forward messages
to eliminate one extra message). Finally, in random
\systemName{}, the number of concurrency control threads that a single
transaction acquires its locks from is, on average, a factor of 3
higher than 
in dual \systemName{}.  The difference in throughput of
each of these systems is therefore directly attributable to the
difference in the number of message delays required for lock
acquisition.


\figref{ycsb_readonly_low} also shows that both \locking{} and 2PL
outperform random \systemName{}. The reason is that messaging overhead
in random \systemName{} exceeds the overhead of uncontended latch
acquisition and data movement in the 2PL and deadlock-free
baselines. However, single and dual partition outperform both 2PL and
deadlock-free locking, indicating that the reduction in
synchronization and data movement overhead does indeed pay off, even
under low contention.

\figref{ycsb_readonly_high} compares each of these baselines under
high contention. 
Since read-only transactions never
conflict with one another, the increase in contention
does not affect the number of actual conflicts in the workload (which
remains zero).
In this case, the throughput of each \systemName{}
configuration increases slightly. The increase in throughput is due to
better cache locality; both concurrency control and execution threads
experience better cache locality. 
Concurrency control threads experience better
locality because they often update the same meta-data across
different transactions (corresponding to contended records). 
Execution threads experience better locality because they read
the same contended records across transactions.


However, unlike \systemName{}, \figref{ycsb_readonly_high} indicates that
\locking{} and 2PL does not scale beyond 60
cores. Instead, the
throughput of these baselines \textit{decreases} after 60 cores. The
decrease in throughput occurs despite the absence of conflicts among
read-only transactions. The reason that these baselines 
does not scale is that multiple transactions requesting read locks on
the same records directly translates to
contention for memory words. In order to acquire or release a logical
lock, both locking implementations must acquire a latch. This latch
protects the hash-bucket in which requests for a particular lock
reside. 
Under high contention, database threads will contend for the latches
protecting the meta-data corresponding to popular
records. Furthermore, as the number of cores allocated to the database
increases, latch contention increases because more threads attempt to
acquire the same set of latches. 
As \secref{conflated} explained, contention on memory words -- such as
those corresponding to latches -- 
leads to deleterious cache coherence overhead,
which in turn inhibits 
\locking{} and 2PL's scalability\footnote{Note
that while our implementation uses latches, latch-free algorithms are
subject to the same cache-coherence overhead
\cite{faleiro2015multiversion, tu2013speedy}.}. Note that \locking{}'s
throughput is nearly identical to that of 2PL. This is because 2PL's
deadlock handling logic is never invoked due to the absence of logical
conflicts among read-only transactions.  \\ \\

\subsection{10RMW transactions}
\label{sec:ycsb_10rw}

\begin{figure} \centering
\includegraphics[width=\linewidth]{\figdir{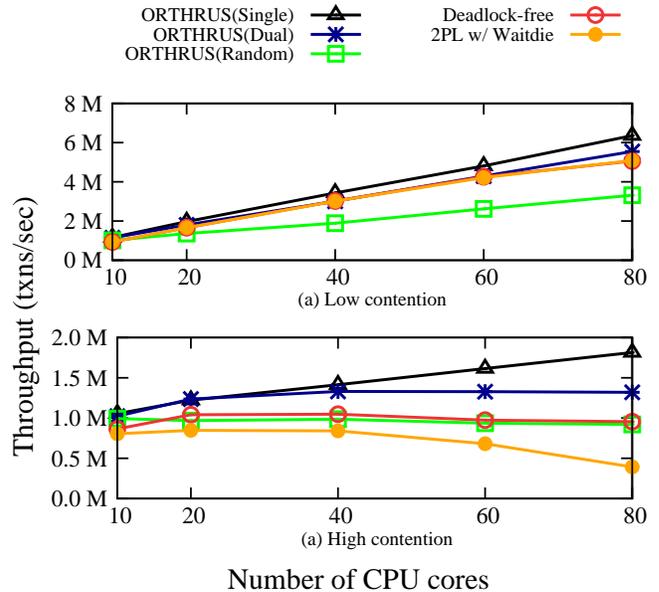}}
\subfigure{\label{fig:ycsb_10rmw_low}}
\subfigure{\label{fig:ycsb_10rmw_high}}
\vspace{-2em}
\caption{YCSB 10RMW scalability. \textbf{(a)} Low contention. 
\textbf{(b)} High contention.}
\label{fig:ycsb_10rmw}
\end{figure}

This section compares \systemName{}'s throughput with that of \locking{}
and 2PL under a workload consisting of transactions that perform 10
read-modify-write operations. 
As in the read-only experiment (\secref{ycsb_readonly}), we
perform a low contention and high contention experiment. 
Records in transactions's read- and write-sets are picked in the same
manner as the read-only experiment. 
In the low contention experiment, all 10 records in a transaction's
read and write-sets
are selected uniformly at random from a set of 10,000,000 records.
In the high contention experiment, each transaction picks 
two records from a set of 64 hot records, and the remaining eight records
from the remaining records (hot records are updated before cold
records). 
Note that
unlike read-only transactions which never logically conflict, a pair
of 10RMW transactions conflict with each other if the records in their
read- and write-sets intersect.


\figref{ycsb_10rmw_low} shows the results of the low contention
experiment. Each system's performance trend is similar to
that in the low contention read-only experiment. 
However, the absolute throughput of
each system is less than its throughput in the low
contention read-only experiment.
This is because 10RMW transactions are longer than read-only
transactions; each 10RMW transaction performs 10 updates, unlike
read-only transactions, which read, but do not update, records.

\figref{ycsb_10rmw_high} shows the throughput of each system under
high contention. Under high contention, 2PL does not
scale beyond 20 cores. On the contrary, its throughput begins to
\textit{drop} as more cores are added. \locking{}'s
throughput also does not scale beyond 20 cores, however, unlike the
dreadlock implementation, its throughput plateaus at 1,000,000
transactions per second. We attribute the difference between
\locking{} and 2PL to deadlock handling overhead. 
Our 2PL baseline uses the wait-die deadlock avoidance algorithm, which
aborts a transaction if it requests a lock held by an older
transaction (\secref{eval}). Therefore, in addition to suffering from synchronization
and data-movement overheads associated with lock acquisition, 2PL
suffers additional overhead due to deadlock handling. As a
consequence, 2PL is outperformed by \locking{}.

As expected, \systemName{}'s single and dual partition configurations
outperform the random configuration. This is because
transactions in single and dual partition \systemName{} request locks
from fewer concurrency control threads (relative to random 
\systemName{}), and therefore hold contended locks for shorter
durations (\secref{optimizations}). \systemName{}'s single partition
configuration  
outperforms the dual partition configuration for the same reason. 

%


At 80 cores, the difference between 2PL and random, dual, and single
partition configurations of \systemName{} is 2.3x, 3.35x, and 4.65x
respectively. In addition, \systemName{}'s dual and single partition
configurations outperform deadlock free 2PL by 38\% and 90\%
respectively. 


\end{appendix}

\end{document}